# Modeling the Dynamics of the Coronavirus SARS-CoV-2 Pandemic using Modified SIR Model with the "Damped-Oscillator" Dynamics of the Effective Reproduction Number

*Anne Ginzburg[1,2], Valeriy Ginzburg[2,3], Julia Ginzburg[3], Ana Garcia Arias[4], and Leela Rakesh[4*]*

[1]H. H. Dow High School, Midland MI 48640

[2]Michigan State University, Lansing MI 48824

[3]VVG Physics Consulting LLC, Lansing MI 48619

[4]Department of Mathematics, Center for Applied Mathematics and Polymer Fluid Dynamics, Central Michigan University, Mt. Pleasant, Michigan 48859

[*]*Corresponding author. Email: rakesh1@cmich.edu*

# ABSTRACT

The COVID-19 pandemic has been – and remains, despite all the medical advances –a great catastrophe that upended human lives and caused millions of deaths all over the world. The rapid spread of the virus, with its early-stage exponential growth and subsequent "waves", caught many medical professionals and decision-makers unprepared. Even though epidemiological models have been known for almost a century (since the "Spanish Influenza" pandemic of 1918-20), the real-life spread of the SARS-CoV-2 virus often confounded the modelers. While the general framework of epidemiological models like SEIR (susceptible-exposed-infected-recovered) or SIR (susceptible-exposed-infected) was not in question, the behavior of model parameters turned out to be unpredictable and complicated. In particular, while the "basic" reproduction number, $R_0$, can be considered a constant (for the original SARS-CoV-2 virus, prior to the emergence of variants, $R_0 \sim 2.5$—$3.0$), the "effective" reproduction number, $R(t)$, was a complex function of time, influenced by human behavior in response to the pandemic (e.g., masking, lockdowns, transition to remote work, etc.) To better understand these phenomena, we model the first year of the pandemic (February 2020 – February 2021) for a number of localities (fifty US states, as well as several countries) using a simple SIR model. We show that the evolution of the pandemic can be described quite successfully by assuming that $R(t)$ behaves in a "viscoelastic" manner, as a sum of two or three "damped oscillators" with different natural frequencies and damping coefficients. These oscillators likely correspond to different sub-populations having different reactions to proposed mitigation measures. The proposed approach can offer future data modelers new ways to fit the reproduction number evolution with time (as compared to the purely data-driven approaches most prevalent today).

# 1. Introduction

Recent development of pandemic infectious disease occurrences around the world including Covid-19, SARS, Ebola, N1H1, and Zika have ignited a worldwide interest in thoughtful understanding the blowout disease [1-23]. Modeling and simulation have become the decision-making tool for the 21$^{st}$ century due to enormous computing power of computers and digital software, in order have a check and balance for human and animal diseases [15-18, 23-36]. As we all know that each has its own characteristics and spread-ability and hence model need to be accommodative in such scenario. Such biological confrontation is becoming valid as based on the current Corona virus situation along with its variability.

This illness has claimed many millions of deaths and causalities all over the worlds since its start in January 2020. As the society began to respond to the pandemic, it was necessary to monitor and predict the spread of the disease and develop counter-measures. Public health officials, medical practitioners and policymakers have been relying on a quantitative understanding of infectious disease dynamics using computer simulation and modeling. Many classical pandemic models have been used in real-life situations and many more models have been updated or re-formulated [6,16,27,37-41]. Their sophistication and fundamental nature permit us to expand to consider more and more challenging scenarios that approximate real-world situations.

There are multiple mathematical models of pandemic spread in the literature [1,4,8-12,16-25,40] that have been developed since early 1900s, but the current COVID-19 creates many new challenges for their application, such as: (i)the effect of unreported infected people in the reported case of the COVID in comparison to perceived cases over the entire infected incidence, (ii) the effect of various hygienic/infectiousness surroundings of hospitalized patients, (iii) the assessment of the requirements of hospitals beds, increase in hospital staff and many more. On the positive side, the governments responded quickly with monitoring and compiling various data, and multiple websites have made the data analysis and modeling more straightforward [1-10,20-29,35-55].

We now turned our attention specifically to mechanistic mean-field models of pandemic spread. In particular, the classical susceptible-infectious-removed (SIR) model was developed in the early XX century and utilized to model the pandemics of SARS, H1N1 and MERs, Bird flu etc. [1-5,10-15,40-43]. The main assumption was to divide the total population into three compartmental groups such as the susceptible, $S$, infectious, $I$, and recovered, $R$, assuming the flow is forward-moving flow of infection from human to human [1-8,19,25,44-50]. Original Kermack-McKendrick SIR model [1-10,18-28,40-49] has three coupled ordinary differential equations (ODE) to model the evolution of S, I, and R sub-populations over time. The susceptible people become infected, develop illness, and subsequently recover (or die from the disease). Those who recover develop immunity and thus are no longer susceptible. Eventually the epidemic slows down and then stops when the susceptible population disappears (with most population having developed immunity through either illness and recovery or vaccination) [16, 18, 47,51-55].

SIR and its subsequent modification SEIR (Susceptible-Exposed-Infected-Recovered) models are now widely used as the standard framework to model the COVID-19 pandemic [3-5,18, 20-30,17,25,37,45, 50-56]]. The "standard" versions of these models utilize constant transmission and removal rates. However, this is a simplification that works poorly for the current pandemic, given that the transmission rates are strongly affected by multiple factors, from virus mutations to public health measures like lockdowns, universal masking, and ultimately vaccination [29,38-42,51].

Recently many researchers [25–29, 45,50,56] used time-dependent SIR models altered to the dynamical epidemics for time dependency, which may incorporate underreporting, false positive, and many other uncertainty of estimates. Some also used Poisson models naturally fit count data [28,36-41]. Several works [17-19,29-39, 53-59] used Poisson distributions to model $I$ and $R$ from frequency including Bayesian perspectives by using independent values for transmission and removal rates, still time dependency of some of these parameters is still under debate. Some researchers proposed to incorporate Poisson model to approximate the dynamics transmission and removal rates, and estimate the movements of the epidemic around the globe by predicting number of the infected and recovered people at certain time period for each location and forecast when the curve flattens using the existing tendency of the data.

The two critical challenges in formulating a predictive SIR-type model for a pandemic are as follows. First, one needs to select the right data to work with. The reported "number of positive cases" usually undercounts the true number of infections, often by as much as an order of magnitude, given that most asymptomatic or mild cases never get tested. Thus, it is generally assumed that the number of reported deaths is a more reliable metric. Knowing the number of daily deaths, $d(t)$, the approximate time between infection and death, and the approximate infection fatality rate (IFR), one can back-calculate the rate of infection as a function of time (although it is important to note that IFR itself can change with time). Second, the model parameters depend crucially on the behavior of the population (the use of masks, lockdowns, etc.) Once the news of the pandemic spreads, the behavioral changes lead to the reduction in the number of human contacts and thus, reduction in the rate of virus transmission. This means that the effective reproduction rate, $R(t)$, is reduced. (The effective reproduction rate is the number of newly infected people for each currently infected person; when $R(t) > 1$, the epidemic is spreading, while when $R(t) < 1$, it is self-extinguishing). When, on the other hand, the level of community spread decreases, the population attempts to "return to normality", increasing the number of person-to-person contacts and driving $R(t)$ back up. Over the course of summer and fall of 2020, this resulted in the second and third waves of pandemic in many US states. These effects were incorporated into several data-driven models, e.g., the one at the Institute for Health Metrics and Evaluation (IHME) at the University of Washington [32].

In this study, we propose a new mechanistic approach aimed at capturing the dynamics of $R(t)$. To do that, we use the technique often employed in physics (e.g., in rheology) known as Prony series fitting, and describe $R(t)$ as the sum of two or three damped oscillators, representing different population subsets. We demonstrate that such an approach successfully describes the data for all localities studied here, even on a mean-field (no spatial variations) level. We also performed Monte Carlo simulations to investigate the role of fluctuations and to quantify the model uncertainty.

## 2. The Model

### *2.1. The Susceptible-Infected-Recovered (SIR) Model Equations*

Our starting point is the SIR model described as,

$$S[k] = N_0 - I[k] - SS[k] - D[k] - R[k] - V[k]$$

$$I[k+1] = I[k] + i[k]$$

$$SS[k+1] = SS[k] + ss[k]$$

$$R[k+1] = R[k] + r[k]$$

$$D[k+1] = D[k] + d[k]$$

$$V[k+1] = V[k] + v[k]$$

(1)

Here, *k* is the day at which the data are recorded (k = 1 is the start of the epidemic in the particular locality), $N_0$ is the total population, *S[k], I[k], R[k]* are the total numbers of susceptible, infected, and recovered, while *SS[k], D[k],* and *V[k]* are the total numbers of sick, dead, and vaccinated. In addition, *i[k], r[k], ss[k], d[k],* and *v[k]* represent the daily changes in the numbers of infected, recovered, sick, dead, and vaccinated. (As usual, to smooth out the fluctuations in the data collection, we compare these with the reported 7-day averages).

Equations (1) are written in a form of recursive relations (RR), rather than ordinary differential equations (ODE) used in various infectious disease modeling [16-23,31-35,45-48]. In principle, the results should be the same independent whether the RR or ODE form is used – the RR form is essentially a discretization of the ODEs where the time increment *dt* = 1 day. Below, we will use RR and ODE descriptions interchangeably.

The daily changes in the numbers of various sub-groups are described by,

$$i[k] = Zp_I I[k]\left(\frac{S[k]}{N_0}\right) - p_s I[k] - p_{ir} I[k]$$

$$ss[k] = p_s I[k] - p_r SS[k] - p_d SS[k]$$

$$r[k] = p_{ir} I[k] + p_r SS[k]$$

$$d[k] = p_d SS[k]$$

$$v[k] = v_0 \left(1 - \exp\left\{-\frac{k - k_0}{\tau}\right\}\right)$$

(2)

In writing equation 2, we made an assumption that most infections come from those in the infectious (I) population; the transmission of the virus from those already in the sick (SS) group is relatively small. The reason for this is two-fold: (i) most infected people are asymptomatic and never develop full illness but might still transmit the virus to others; (b) people who are already sick have fewer contacts and those around them typically use protective equipment making the transmission less likely.

The model parameters are regressed based on the early-stage COVID-19 data and are summarized in Table 1. (The parameters related to vaccination, $v_0$, $k_0$, and $\tau$ are not included in the table and will be described later).

*Table 1. SIR Model Parameters*

| |
|---|
| $p_i = 0.15 (/day)$ |
| $p_s = 0.05 (/day)$ |
| $p_{ir} = 0.2 (/day)$ |
| $p_r = 0.05 (/day)$ |
| $p_d = 0.002 (/day)$ |
| $Z_0 = 5$ |

Based on the model equations, we can define two crucial parameters, the effective reproduction number, *R*, and the infection fatality rate, IFR,

$$R = \frac{Zp_i}{p_s + p_{ir}} \qquad (3a)$$

$$IFR = \frac{p_d}{p_s + p_{ir}} = 0.008 \qquad (3b)$$

The number of contacts per person per day, *Z*, is the crucial variable that determines the evolution of the viral spreading. We assumed that at the beginning ("normal state"), $Z = Z_0 = 5$, and $R = R_0 = \frac{Z_0 p_i}{p_s + p_{ir}} = 3.0$. Subsequently, *Z* becomes affected by measures like lockdowns. In principle, the probability of transmitting the virus, $p_i$, also depends on the human behavior (e.g., masks can reduce $p_i$ by 30—70%). However, given that *Z* and $p_i$ enter into the model only as the product $Zp_i$, we can combine these two effects together and just analyze the dependence of $Zp_i$ or *R* on time.

### 2.2. *Equations of Evolution for the Reproduction Number*

Let us begin by analyzing what the data are telling us. To do that, we allow *R (*or *Z)* to float while keeping all the other model parameters constant. We then calculate the cumulative death number *D[k]* and manually fit it to the data from COVID Tracking Project [ 10-15,20,29, 51,55-66] . (Once again, using the death, rather than positive test, data is considered to be a more reliable method of analysis). Using the data for the State of Michigan, we back-calculated the following evolution of the effective reproduction number, *R(t)*, as shown in Figure 1 (where Day 1 is 02/15/2020, when the first COVID-19 case was registered in the State of Michigan).

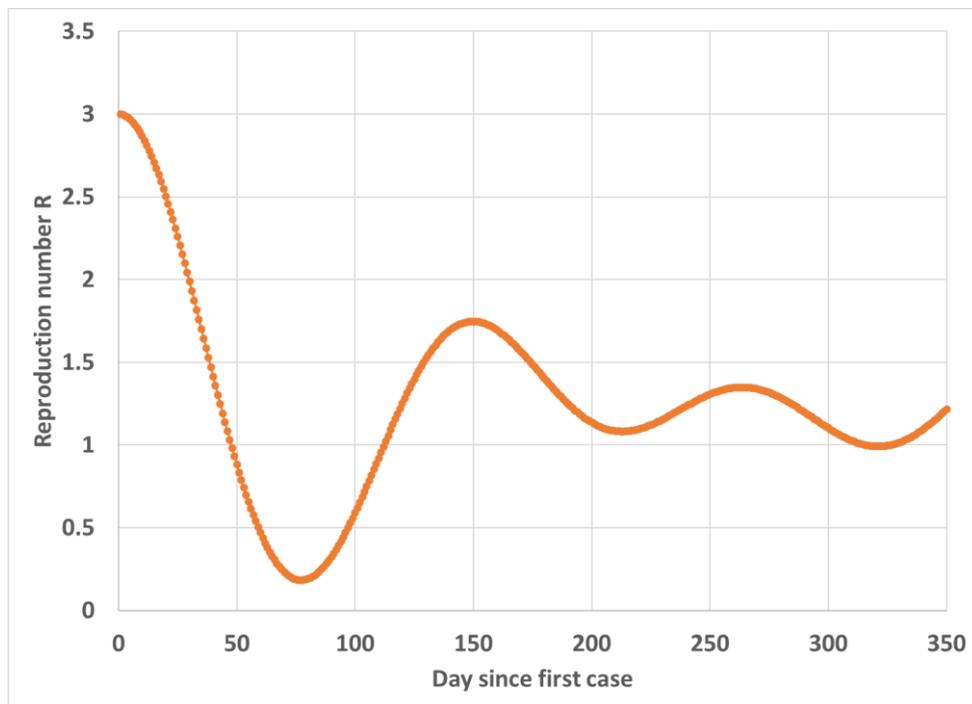

*Figure 1. Estimated reproduction number, R(t), as a function of time, for Michigan.*

We notice that the curve in Figure 1 bears a striking resemblance to the behavior of "damped oscillators", like viscoelastic elements with inertia in fluid dynamics (rheology) or Inductor-Capacitor-Resistor (LCR) circuits in electrical engineering. To describe the process quantitatively, we thus make two assumptions. First, we assume that the overall reproduction number *R(t)* comes from averaging over different sub-populations (see below for more detailed discussion), $R = \sum_{l=1}^{m} w_l R_l$. Second, we propose to describe the evolution of $R_l(t)$ using the following differential equation model:

$$\frac{d^2 R_l}{dt^2} + \Gamma_l \frac{dR_l}{dt} + \omega_{0,l}^2 \left( R_l - R^* \right) = 0 \qquad (4)$$

Here, the index $l$ corresponds to different sub-populations; $\Gamma_l$ is the dissipation term, and $\omega_{0,l}$ is the natural frequency of the oscillator. The initial conditions are as follows, $R_l(0) = R_0$ and $R_l'(0) = 0$ (where $R_l'(t) \equiv \frac{dR_l(t)}{dt}$). The resulting solution is then written as,

$$R = \sum_{l=1}^{m} w_l \left[ R^* + (R_0 - R^*)\exp\{-\gamma_l t\}\left\{\cos(\omega_l t) + \frac{\gamma_l}{\omega_l}\sin(\omega_l t)\right\} \right] \quad (5)$$

In principle, $\gamma_l$ and $\omega_l$ should be expressed as functions of $\Gamma_l$ and $\omega_{0,l}$; however, given that none of these parameters can be derived from first principles, we will work directly with $\gamma_l$ and $\omega_l$ and optimize them in our modeling. The summation over $l$ corresponds to the summation over sub-populations, $m$ is the number of sub-populations, and $w_l$ is the "weight" of each sub-population. We consider up to three sub-populations within each "closed locality" (either US state or a country).

The parameters in equation 5 are then determined based on the following procedure. The COVID-19 death data for a locality are compiled, and $R(t)$ is back-calculated. We then use Excel Generalized Reduced Gradient (GRG) method to minimize the following objective function,

$$\chi = \frac{1}{N}\sum_{t=1}^{N}\left|\ln(R_{\exp}(t)) - \ln(R_{\text{model}}(t))\right| \quad (6)$$

Here, $R_{\exp}$ is the regressed reproduction number (e.g., the curve in Figure 1), and $R_{\text{model}}$ is given by equation 5. We perform the GRG minimization several times, using different initial trial parameters, until the objective function of equation 6 becomes sufficiently small enough.

In Figure 2, we show the comparison of model and data for the COVID-19 death data in Michigan. It can be seen that the model successfully reproduces the data. Only two sub-populations ($m = 2$) were necessary to achieve this agreement.

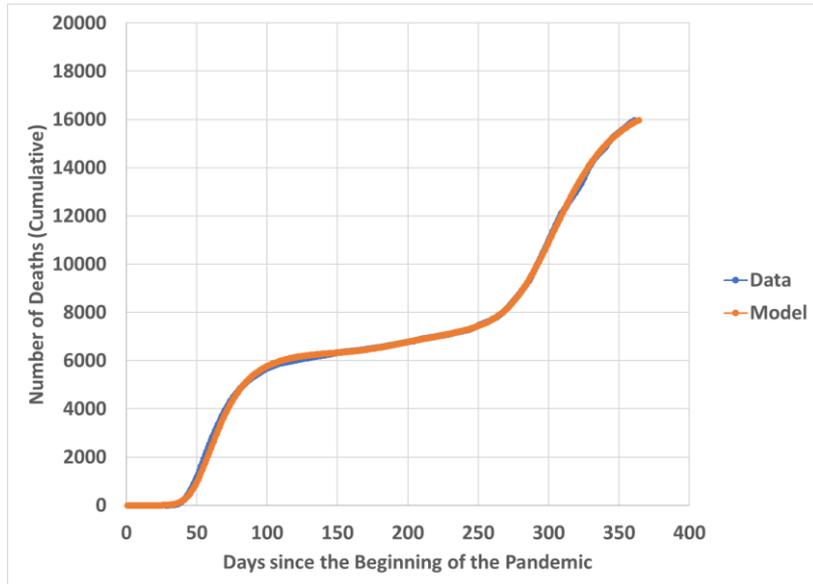

*Figure 2. Recorded deaths (blue) and model fit (orange) for COVID-19 in Michigan during the first year (2/15/2020 – 2/14/2021).*

The effective reproduction numbers for the two sub-populations are plotted in Figure 3. Obviously, for both sub-populations, R(t) has the standard form of decaying oscillations. The first sub-population (represented by the grey curve) seems to have responded to the pandemic in a very forceful way, by rapidly curtailing its R-number to near-zero in the first two months. This probably corresponds to significant masking, adherence to staying at home most of the time, etc. This sub-population also seems to recover to pre-pandemic patterns of behavior once the numbers go down (days 100 – 150). The second sub-population (blue curve) exhibits similar variations, but they are less dramatic and occur over longer timeframes.

We are reluctant to ascribe specific meaning to these sub-populations – or even whether their composition changes with time or stays approximately the same. This is more the task for sociologists and public health experts – our goal is primarily to provide a mathematical formalism to develop these models.

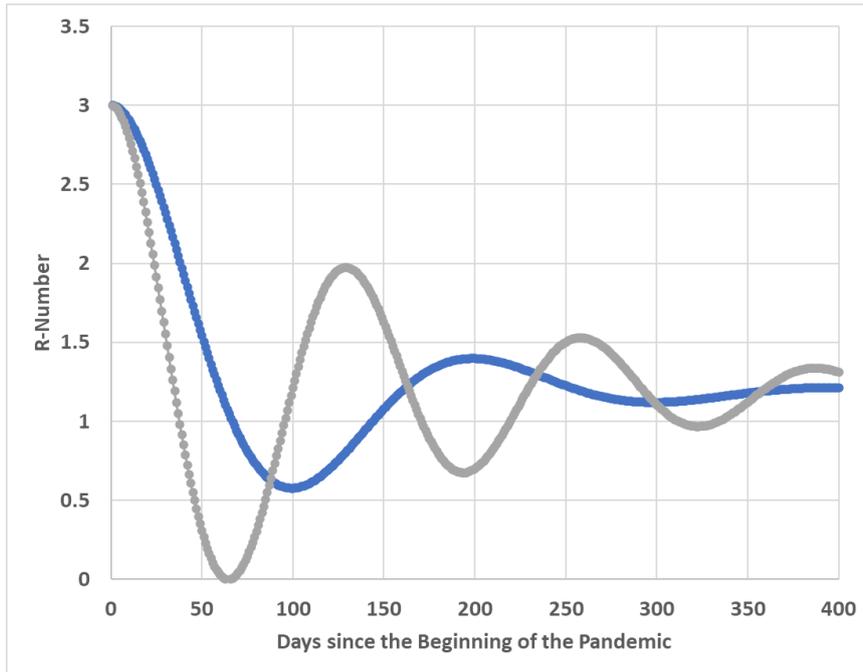

*Figure 3. Reproduction number, R(t), for two sub-populations in Michigan, based on the model.*

Once the model is parameterized (based on the death data), we can back-calculate other characteristics in the SIR model as shown in Figure 4.

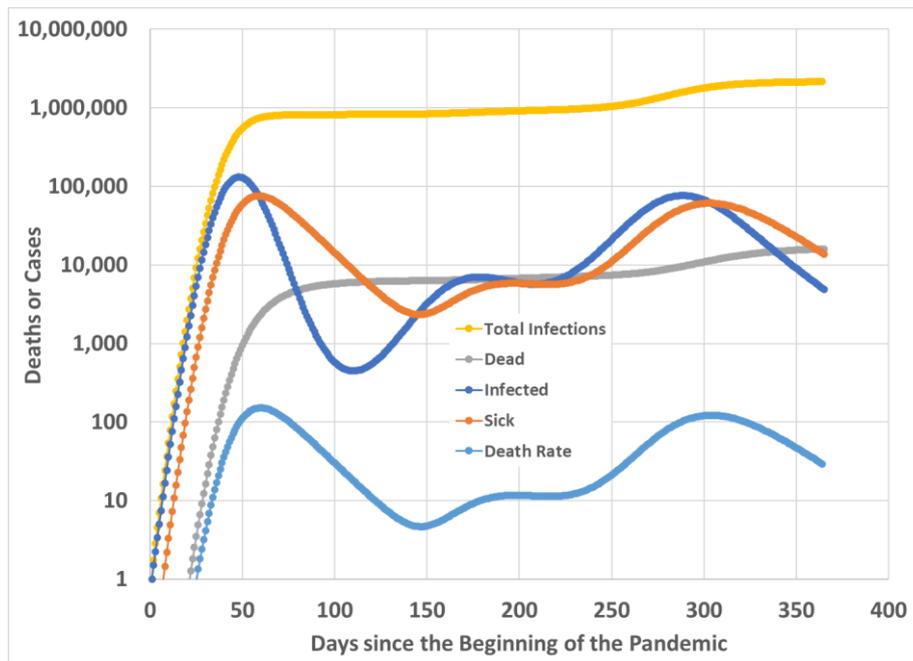

*Figure 4. Model-estimated total infections (yellow), 7-day averaged new infections (dark blue), sick or symptomatic (orange), and 7-day averaged number of new deaths (light blue) for Michigan during the first year of pandemic. Note that the Y-axis is log-scale.*

The model captures the main features of the last year (2020) in Michigan:

- Sharp initial peak between Day 20 and Day 100 (early March to late May 2020);
- Weak peak in summer of 2020;
- Strong and broad peak in winter between Days 250 and 350 (late October 2020 to early February, 2021)

We estimated that by February 2021, when this study was concluded, about 20% of the population had been infected at some point. About half of those infections (~1MM people) took place during the winter spike, from late October 2020 to early February 2021. We also estimated that the number of people infected by the virus was about five times greater than the reported number, due to the lack of testing at the early stages of the pandemic and the large number of asymptomatic infections. This estimate is consistent with the literature.

The model allowed us to make predictions, as shown in Figure 5. The predictions were made for four months into the future (February – June 2021). The assumptions were as follows:

- The nature of the virus (specifically, its basic reproduction number $R_0$) does not change.
- The immunity does not wane with time.
- The R(t) pattern continues without changes.
- The vaccination campaign can be described by equation, $v[k] = v_0 \left(1 - \exp\left\{-\frac{k-k_0}{\tau}\right\}\right)$

  Vaccination starts on Day $k_0 = 320$ (December 31, 2020) and ramps up over $\tau = 120$ days. Maximum vaccinations per day capped at 50k (based on 1.8 million/day nationwide and assuming proportional allocation between states). Vaccination stops when 50% of the population is vaccinated. Everyone is considered fully immune after the first shot.

We then estimated the number of deaths by June 1, 2021 and compared those with the predictions from IHME (more elaborate data-driven model). This analysis for Michigan and other states will be given in Section 3.

## 2.3. Incorporation of Stochasticity into the SIR Model

The analysis above is "simple mean-field", with no accounting for fluctuations. In order to estimate potential error bars for the model, it is necessary to build-in some measure of stochasticity and randomness. To do this, we adopt the following procedure. For the "past" (February 2020 – February 2021), the reproduction number R(t) is calculated as described in the previous section. For the "prediction" part (March – June 2021), calculations are repeated ten times, with the addition of Gaussian random noise,

$$R(t) = R_{model}(t) + A(ran(t) - 0.5). \qquad (7)$$

Here, $R_{model}$ is given by equation 5, *ran(t)* is the (pseudo)-random number with equal distribution between 0 and 1, and A is the fluctuation amplitude (here set to A = 0.1). For each run with the pseudo-random variation in R(t), we calculate the death numbers, and then analyze those results to obtain the average, the standard deviation, and the percentiles. For Michigan, these results are shown in Figure 5 and Table 2.

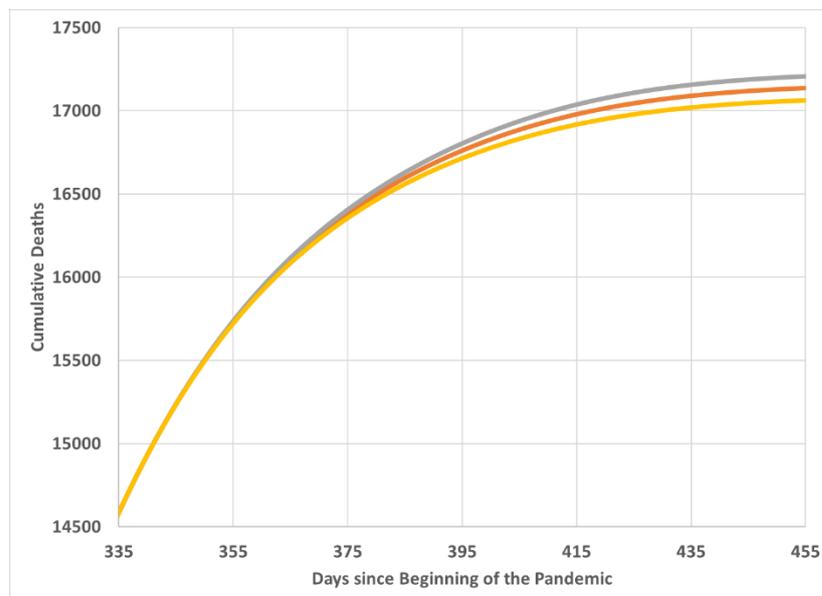

*Figure 5. Predictions for cumulative COVID-19 deaths for March – June 2021: (orange) average, (grey) upper bound, and (yellow) lower bound.*

*Table 2. Calculated 5%, 25%, 50%, 75%, and 95% percentiles for cumulative COVID-19 deaths in Michigan for 6/1/2021.*

| Percentile | Calc Deaths |
|---|---|
| 5% | 17,130 |
| 25% | 17,143 |
| 50% | 17,165 |
| 75% | 17,179 |
| 95% | 17,197 |

Note that these predictions were made in February 2021 and did not account for the emergence of the Delta variant (and subsequently Omicron variant) with significantly higher $R_0$. However, the impact of Delta was not felt until later in the summer of 2021, so the model calculations described here do not diverge from the actual data very significantly (<20%).

The analysis of averages, standard deviations, and percentiles for several selected states and countries is provided in the next section.

## 2.4. Monte Carlo Simulations

Finally, we attempt to consider the role of spatial variations, including the formation of "hot spots". Neither the "mean-field" SIR nor the "stochastic" version can capture these phenomena. Instead, we utilized lattice Monte Carlo approach, with the main features as follows.

- Each resident is represented by a point on a lattice.
- Each point can be in one of several states: S (susceptible), I (infected), SS (sickened), R (recovered), V (vaccinated), or D (dead).
- Each infected person can transmit virus to one of their 48 "neighbors".

Transitions between states are determined using Metropolis algorithm, with the probabilities of transitions given in Table 1.

This approach, while cannot be applied to large population areas like states and countries, can be useful to describe smaller communities (cities, counties, universities, schools, etc.) Here, we model the pandemic spread in Midland County, Michigan (population close to 87,000 people). We utilize two-dimensional square lattice 300x300 (translating to 90,000 residents) and running ten Monte Carlo runs. The results will be discussed in the next section.

## 3. Results and Discussion

### *3.1. Deterministic SIR Model – Results for United States*

In Figure 6, we plotted cumulative death numbers for six states, chosen at random out of the fifty. The approach to the fitting is the same as discussed earlier for the Michigan example. In some cases, the model required two sub-populations, while in some others, it was necessary to employ three distinct sub-populations. The same analysis was performed for all other states, and the quality of fit was roughly the same for all of them. Note that the pandemic occurred in different ways in different states. In NY and NJ, the extremely severe early stage of the pandemic was followed by nearly complete suppression of the spread for most of the rest of the year (2020, until the winter 2020-2021 wave. In TX and CA, the first wave was shallower, with prolonged increases and decreases in the death rate, while the second (winter) wave turned out to be significantly worse. In LA, there were at least three waves in the first year, while in AK, discerning any separate waves is difficult altogether. Even so, the approach proposed here seems to have successfully captured all of these diverse behaviors from various states in USA as shown in figure 6.

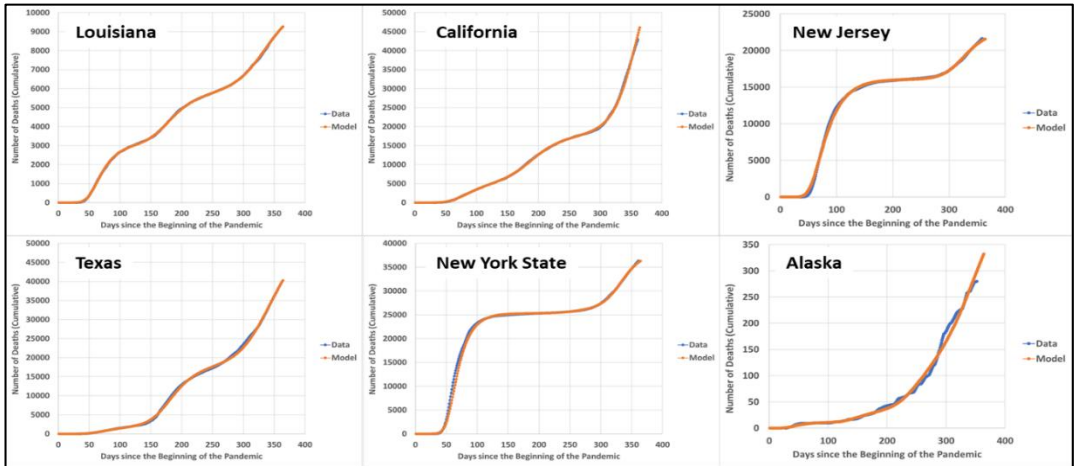

*Figure 6. Recorded cumulative deaths (blue) and model fit (orange) for six US states (LA, CA, NJ, TX, NY, and AK) for the period between 2/15/20 and 2/10/21.*

Figure 7 shows the scatter plot comparing our model with the IHME [32]estimates (pulled from their website on 02/21/21). The comparison was based on the predicted number of COVID-19 related deaths in each of the fifty states as of 06/01/2021. It can be seen that our model compares favorably with the IHME model even while using significantly fewer adjustable parameters. Again, however, the point here is not necessarily to advertise the new approach as the "better" or "more accurate" one, but to show that this "damped oscillator" model captures the essential features of the pandemic and thus could be useful in future pandemics (even when the data-driven models developed for COVID-19 would have to be re-designed completely).

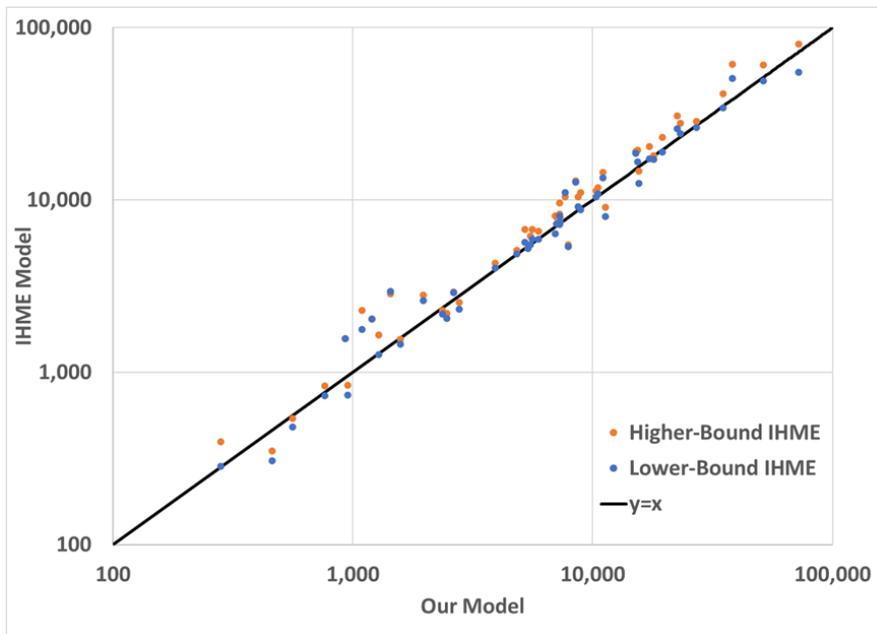

*Figure 7. Scatter plot comparing the cumulative deaths from COVID-19 in fifty states as predicted by IHME compared to our "damped oscillator" model. The predictions were made on 02/21/2021 for 06/01/2021.*

One particular feature of interest in our model is the parameter R* -- the "new normal". This parameter describes the effective reproduction number for the virus at the time when the population has adjusted to the pandemic and no longer modifies its behavior. The value of R* depends on many factors, including population density, availability of services, and willingness of residents to modify their actions (i.e., staying inside, wearing masks, etc.) Figure 8 shows how the estimated R* correlates to the "COVID-19 death rate" (defined as the cumulative number of COVID-19-related deaths per 100,000 residents).

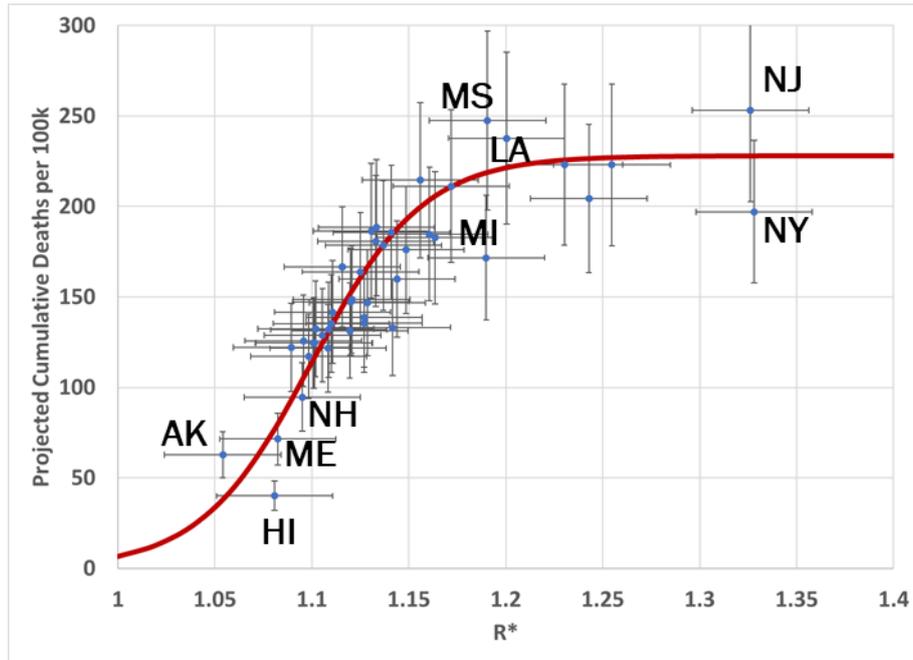

*Figure 8. Projected COVID-19 death rate vs. estimated R\* for the fifty states. The red curve is guide to eye.*

The sigmoidal red line is a guide to eye showing three different regimes:

- R\* < 1.1 corresponds to isolated and/or low-density states (AK, ME, HI) where the pandemic was in the transient collective immunity (TCI) state for much of 2020;
- 1.1 < R\* < 1.2 corresponds to most states where the pandemic had several waves, and the mortality depended on population density, timing of the first wave, and mitigation measures;
- Finally, R\* > 1.2 corresponds to the states where initial losses were very heavy and by the spring of 2021, the herd immunity threshold (HIT) was already exceeded due to combination of infections and vaccinations.

Obviously, this analysis is very preliminary and more data (including those from non-US localities) are needed to use it for predictive purposes. This is a topic for future research.

### 3.2. Stochastic SIR Model [5,22-24,53-55] – Results from Selected Countries

In this subsection, we describe the SIR analysis with added stochastic terms and use it to estimate not just averages, but also the standard deviation and percentiles, as discussed in Section 2.3.

The results for United States (as a whole), Canada, Italy, Brazil, Norway, Sweden, and France are shown in Figures 9—15.

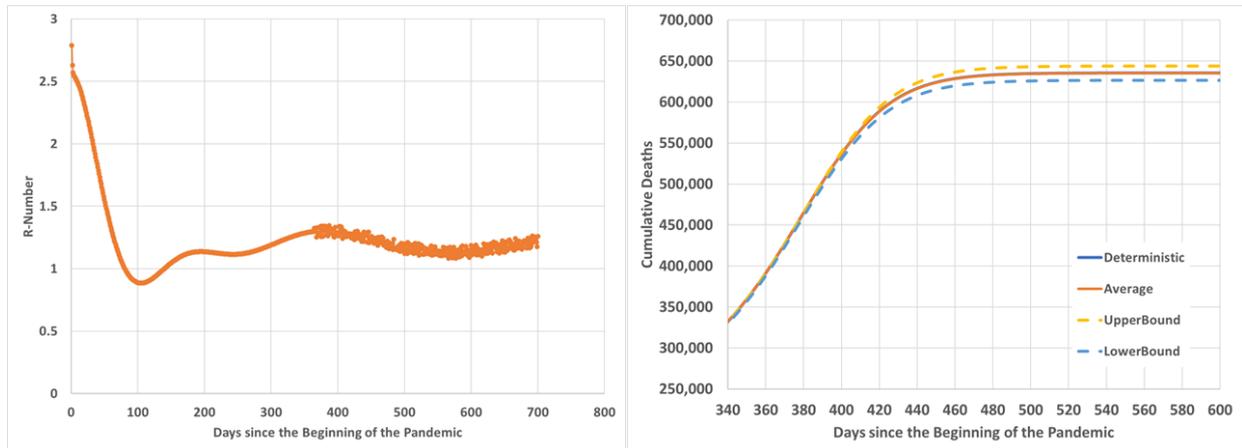

*Figure 9. USA*

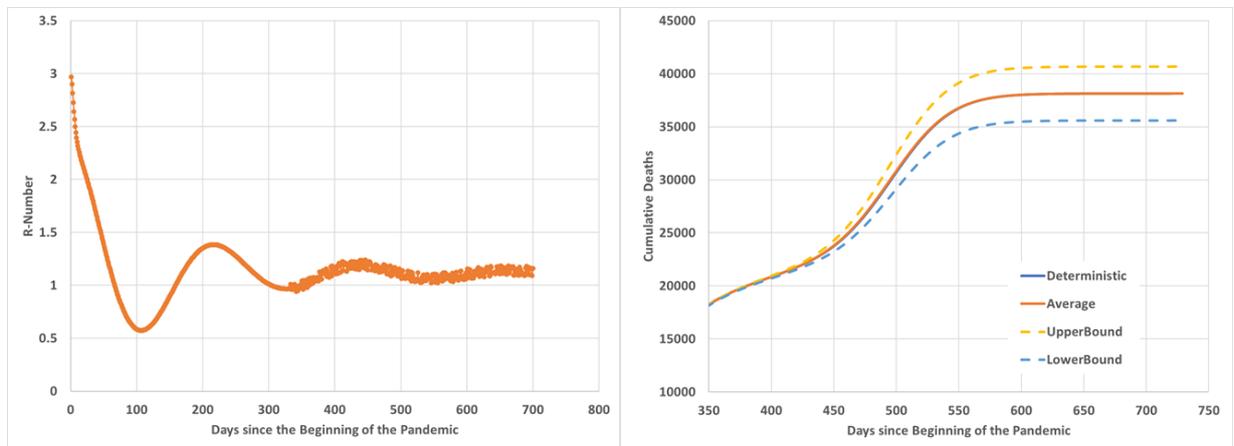

*Figure 10. Canada*

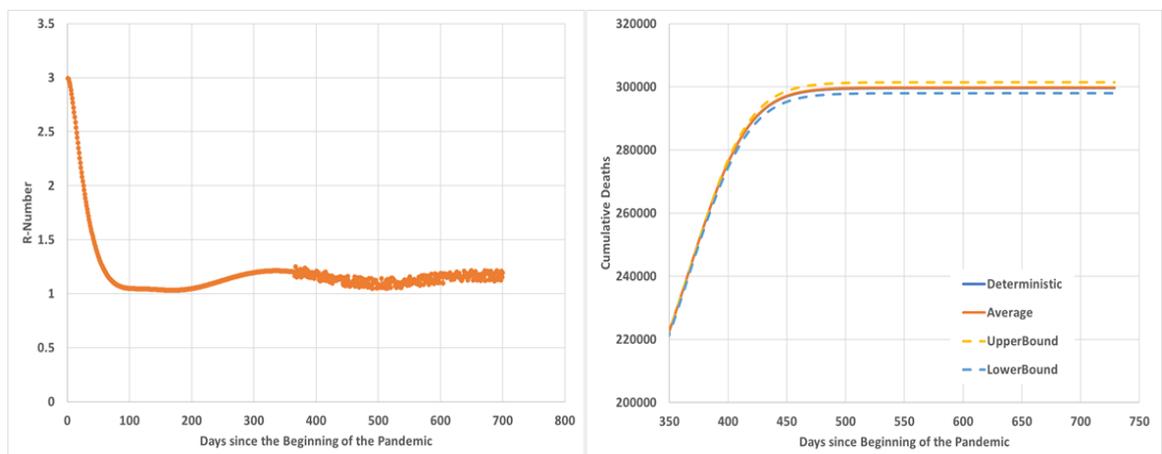

*Figure 11. Brazil*

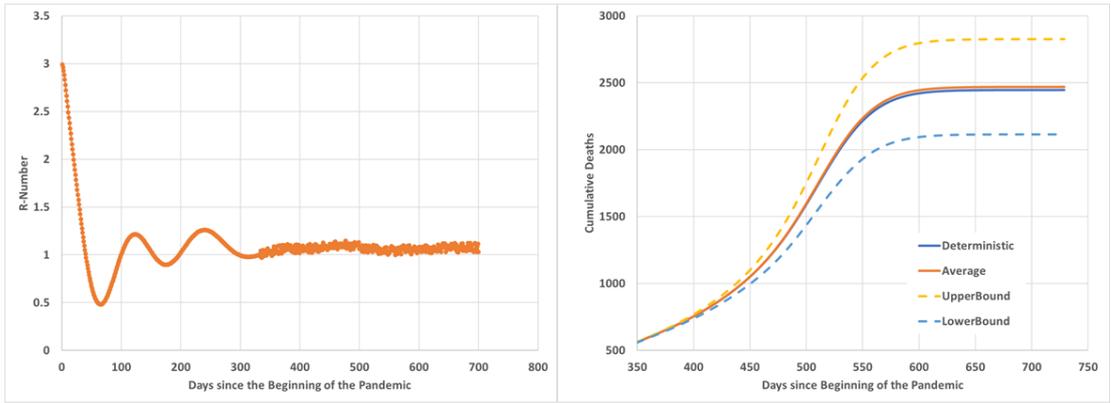
*Figure 12. Norway*

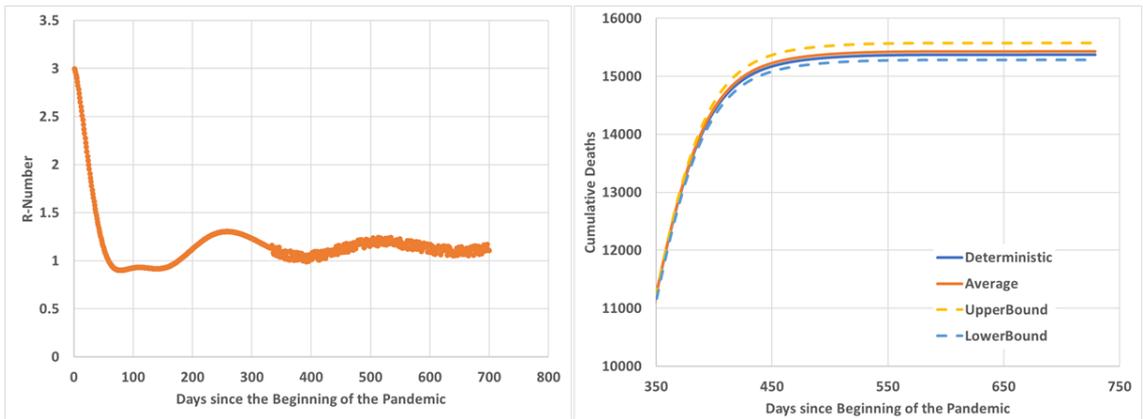
*Figure 13. Sweden*

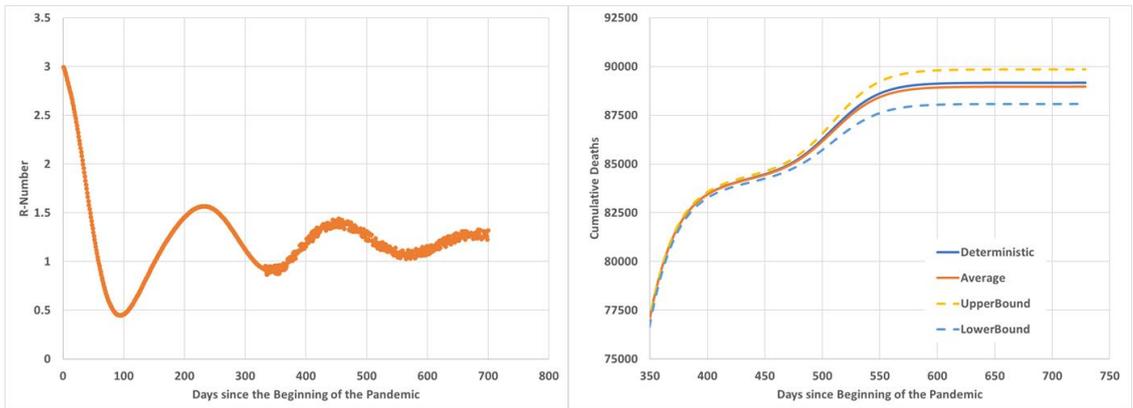
*Figure 14. Italy*

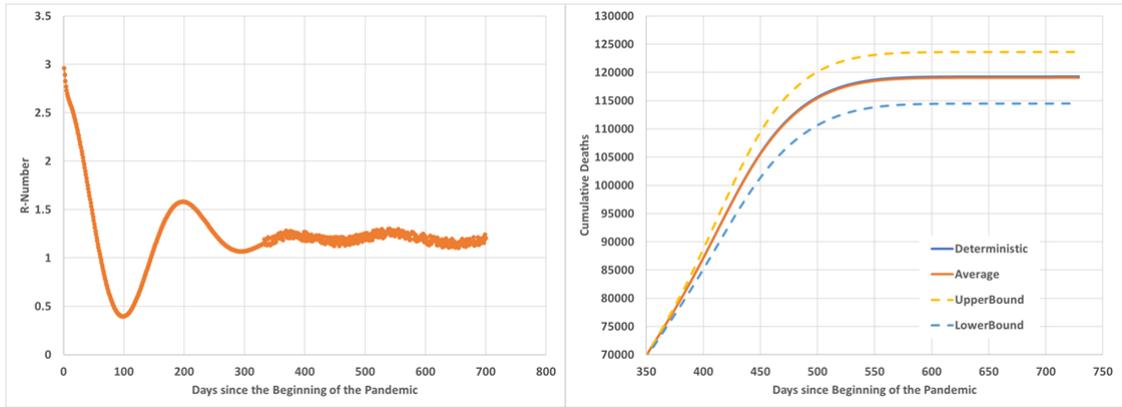

*Figure 15. France*

In all figures above, the left panel represents the best-fit R(t), and the right panel shows the predictions for the cumulative COVID-19-related deaths after March 1, 2021, including the average (orange line), upper and lower bounds (gold and blue dashed lines). The calculated parameters are summarized in Tables 3 and 4.

*Table 3. Calculated R\* and deaths per 100k residents (estimates for 6/1/2021) for seven countries considered in this study*

| Country | R* | Deaths per 100k |
|---|---|---|
| USA | 1.20 | 192 |
| Canada | 1.14 | 150 |
| Brazil | 1.14 | 150 |
| France | 1.20 | 180 |
| Italy | 1.23 | 232 |
| Sweden | 1.14 | 150 |
| Norway | 1.07 | 51 |

*Table 4. Percentile analysis for the seven countries considered in this study*

| Percentile | Calculated Deaths (as of 06/01/2021) | | | | | | |
|---|---|---|---|---|---|---|---|
| | US | Canada | Brazil | France | Italy | Sweden | Norway |
| 5% | 631,433 | 15,384 | 298,888 | 117,050 | 36,869 | 15,384 | 2,283 |
| 25% | 632,742 | 15,402 | 299,348 | 117,540 | 37,825 | 15,402 | 2,408 |
| 50% | 635,682 | 15,415 | 299,645 | 119,523 | 37,957 | 15,415 | 2,492 |
| 75% | 637,400 | 15,448 | 300,262 | 120,277 | 38,656 | 15,448 | 2,547 |
| 95% | 639,713 | 15,513 | 300,458 | 121,043 | 39,411 | 15,513 | 2,617 |

The above analysis shows interesting similarities between various countries and US states analyzed in the previous subsection. For example, the R(t) profile for Italy is very similar to that of Louisiana, while that of France is similar to, e.g., Michigan; there are strong similarities between the R(t) for Sweden and, e.g., North Carolina, and Norway and Hawaii. As suggested by Table 3, there is a strong correlation between the number of deaths per 100,000 residents and the model parameter R* describing the "new normal" for the pandemic behavior. Thus, localities as different as Brazil and Canada, or France, US, and Italy show strong similarities in this regard. Again, we are not attempting to over-analyze these trends, just to show that the pandemic profiles over the first year of COVID-19 were relatively similar in many places.

### 3.3. Monte Carlo Modeling – Results for Midland County, Michigan, USA

Finally, we discuss the effect of spatial heterogeneity, as described by the lattice Monte Carlo modeling. Figure 16 shows several snapshots from a Monte Carlo run for Midland County. Each frame corresponds to a particular time over the course of the pandemic, with the first frame corresponding to 1 week after the first case, and each subsequent frame being 45 days after the previous one. The color coding is as follows – white, Susceptible; yellow, Infected; green, Recovered; red, Dead; teal, Vaccinated. Given that most of Infected persons become Recovered within the 45-day interval between frames, it is primarily the map of Recovered people that shows the spread of the virus in the population. The vaccination is assumed to have started in December 2020, i.e., between the fifth and sixth frames. By the time corresponding to the ninth frame, about 50% of the county population has been vaccinated, according to the model. The spread of the virus in this run effectively stops somewhere between frames 6 and 7 (day 232—277, or mid-winder 2020-2021).

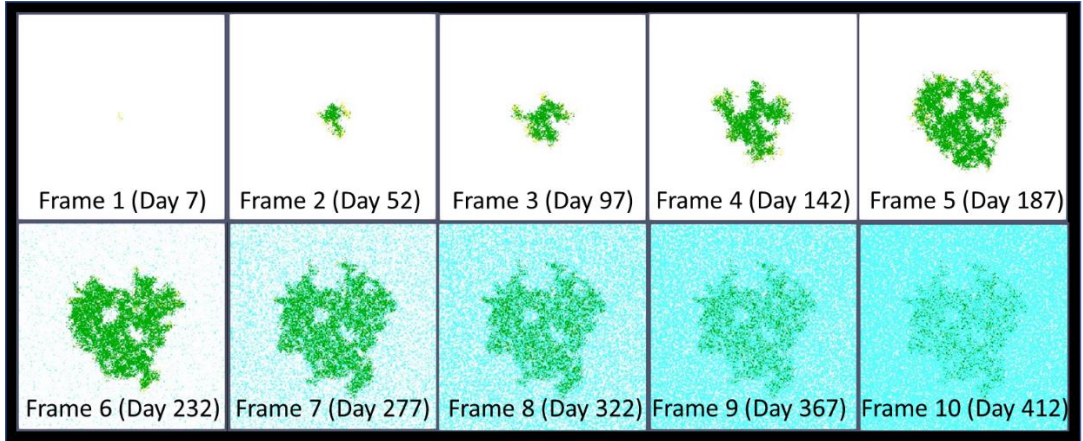

*Figure 16. Monte Carlo run snapshots. See text for more details.*

Of course, different runs exhibit different patterns of spreading, and the numbers of infections and deaths in the community vary depending on the random numbers used to initialize a Monte Carlo run. In Figure 17, we show the average, upper, and lower bounds for ten Monte Carlo runs. The variability is very substantial, with the total number of deaths changing between 40 and 140, with the average of 88, somewhat lower than the deterministic SIR value of 114.

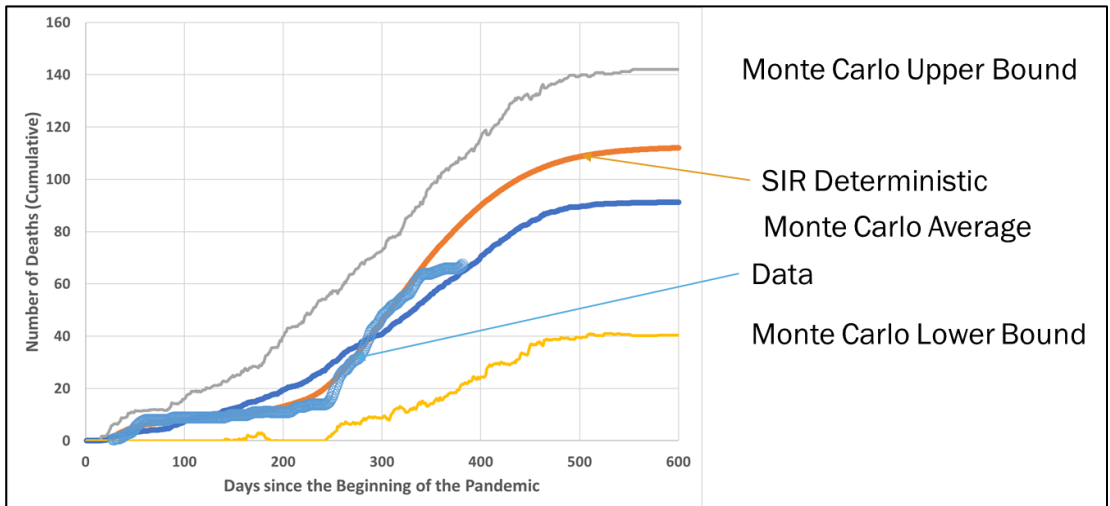

*Figure 17. Monte Carlo simulation for Midland County, Michigan. See text for more details.*

### 3.4. Discussion

The above analysis combines two approaches – the SIR model to describe the mechanism of virus spreading, and the "damped oscillator" model to describe the evolution of the SIR model parameters (in particular, the effective reproduction number, $R(t)$) in response to the behavioral changes caused by the pandemic (such as mandatory quarantines, sheltering in place, universal masking, travel restrictions, etc.) While the former has been well-known and accepted for about a hundred years, the latter is still representing a great challenge that, in principle, requires combined efforts from mathematicians, public health experts, and sociologists. In 2020, fundamental understanding of the relationship between $R(t)$ and various measures introduced by public health authorities was sorely lacking, thus hobbling the search for best mitigating measures.

Most attempts to model the COVID-19 pandemic in 2020 were based on purely data-driven modeling, i.e., machine-learning (ML) and artificial intelligence (AI) analysis of the past trends and their extrapolation into the future. The specific techniques used primarily involved neural networks (NN) and genetic algorithms (GA). The resulting models showed some success in predicting the short- and medium-term future development of the pandemic, but provided very limited mechanistic description of various factors governing the changes in R as a function of time. Given the expected relationship ("feedback loop") between the disease prevalence in the community and the effective reproduction number, one should, at least in principle, strive to uncover more fundamental trends for the evolution of $R(t)$. At the same time, the complexity and nonlinearity of the problem make it difficult to attack it using a purely first-principles approach with a dozen or so adjustable parameters that could be fitted willy-nilly to fit the data.

The approach proposed here, thus, attempts to find a middle ground between a pure data-driven analysis and a fully first-principles-based theory (see Figure 18).

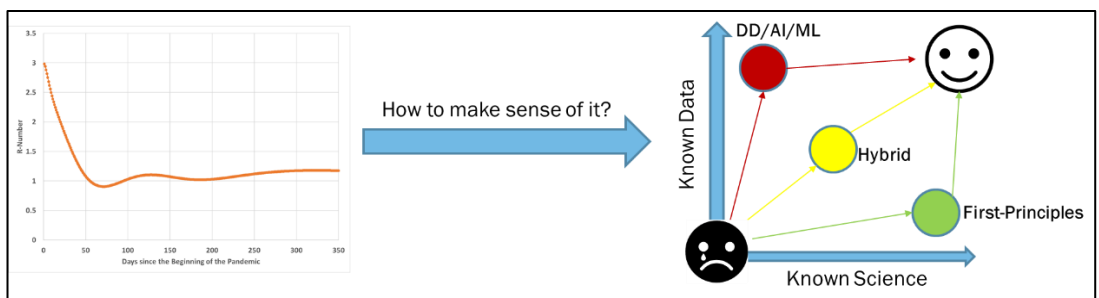

*Figure 18. Schematic representation of various modeling approaches. DD – data driven, AI – artificial intelligence, ML – machine learning. The approach used in this study is characterized as "Hybrid".*

Our model for R(t) is based on two main ideas. First, that the population response to the pandemic is similar to the response of an LCR (inductance, capacitance, and resistance in series) electrical circuit to the external AC voltage – or the response of viscoelastic Maxwell element with inertia to the external oscillatory force. This means describing R(t) – or, alternatively, the effective number of contacts per person per day, Z(t) – as a "damped oscillator". The idea is that the behavior of the population changes from the pre-pandemic normal to some "new normal" where many leisure activities are curtailed and many essential activities are transitioned online. However, the process of finding this "new normal" is not straightforward and goes through some decaying oscillations before reaching its equilibrium. Second, that the population is, in fact, non-uniform, and must be described as a sum of two or more sub-populations. The amplitudes and the periods of oscillations between the pre-pandemic and the pandemic equilibria are different for different sub-populations. We are not making determination as to what these sub-populations are (urban vs. rural? Old vs. young? Essential workers vs. others?), but rather are trying to fit the patterns and determine the relative weights of those sub-populations. It will then be up to sociologists to hypothesize the meaning of these numbers. One particular question in this regard is whether the long-term asymptotic value (R*) is the same or different for different sub-populations (we assumed it to be the same).

Our analysis suggests, rather intriguingly, that the pandemic dynamics in 2020 and early 2021 (before the onset of the Delta variant) were similar in many places that look very different – or, indeed, adopted very different response strategies. At the same time, there still was no universal pattern – some places had significantly higher R* and death rates, compared to others. Understanding these trends will remain a challenge for years to come.

## 4. Conclusions

We showed that a simple SIR model can successfully describe the dynamics nature of a pandemic like COVID-19, provided that the population response (as represented by the time-dependent reproduction number, R(t) is appropriately parameterized. We then proposed that the dependence of R(t) can be captured by assuming that the population response can be modeled as "damped oscillators" with different decay constants and oscillation periods. The combined analysis was applied to all 50 US states and several countries (US, Canada, Brazil, France, Italy, Sweden, and Norway). To the mechanistic mean-field SIR model, we added temporal and spatial randomness to estimate the uncertainties, standard deviations, upper and lower bounds. The model was able to successfully describe the observed data for the first year of the pandemic, prior to the emergence of the Delta variant.

The proposed "damped oscillator" phenomenological approach describing the sub-populations and their dynamics can be tied to sociological studies of human mobility and the "damped oscillator" parameters could be related to specific characteristics of a particular society (rural vs. urban, young vs. elderly, etc.) We expect this will be a subject for future studies.

# References

1. Anesi, G. L., et al. (2021). Characteristics, Outcomes, and Trends of Patients With COVID-19-Related Critical Illness at a Learning Health System in the United States. Annals of internal medicine, 174(5), 613–621. https://doi.org/10.7326/M20-5327

2. Allen, L.,Van den Driessche, P. (2006). Stochastic edipdemic models with backward bifurcation, Mathematical Biosci. 3, 445-458

3. Barry, J. (2009). Pandemics: avoiding the mistakes of 1918. *Nature* 459, 324–325.

4. Bates Ramirez, V. (2020, April 20) What is R. health-line. doi.org/DOI: https://www.healthline.com/health/r-nought-reproduction-number#covid-19-r-0

5. Baxter, Rodney J. (1982). Exactly solved models in statistical mechanics. Academic Press Inc. ISBN 9780120831807.

6. Blackwood, J.C., Childs, L.M. (2018). An introduction to compartmental modeling for the budding infectious disease modeler (5th edition). Letters in Biomathematics. doi.org/DOI: https://doi.org/10.1080/23737867.2018.1509026

7. Boatto, S., Bonnet, C., Cazelles, B., Mazenc, F. (2018). SIR model with time dependent infectivity parameter: approximating the epidemic attractor and the importance of the initial phase. CCSD. https://hal.inria.fr/hal-01677886

8. Brauer, F., Kribs, C. (2015). Dynamical systems for biological modeling: An introduction. Boca Raton. CRC Press.

9. Brauer, F., Van den Driessche, P., Wu, J. (Eds.). (2008). Mathematical epidemiology. Lecture Notes in Mathematics. Vol. 1945. Berlin: Springer.

10. Buchwald, A.G., Adams, J., Bortz, D.M., Carlton, E.J. (2020). Infectious Disease Transmission Models to Predict, Evaluate, and Improve Understanding of COVID-19 Trajectory and Interventions. Annals of the American Thoracic Society, 17(10), 1204–1206. doi.org/DOI: https://doi.org/10.1513/AnnalsATS.202005-501PS

11. Burke, D.S. (2003). Computational modeling and simulation of epidemic infectious diseases. Microbial threats to health: emergence, detection, and response. Washington, DC: National Academies Press. 2003: 335–341. doi.org/DOI: https://www.ncbi.nlm.nih.gov/books/NBK221490/

12. Center of Disease Control and Prevention. (2021). Using an Epi Curve to Determine Mode of Spread. doi.org/DOI: https://www.cdc.gov/training/QuickLearns/epimode/

13. Chandra, S., Kassens-Noor, E. (2014). The evolution of pandemic influenza: evidence from India, 1918–19. BMC Infectious Diseases, 14. doi.org/DOI: https://bmcinfectdis.biomedcentral.com/articles/10.1186/1471-2334-14-510

14. Chen, Y.C., Lu, P.E., Chang, C.S., Liu, T.H. (2020). A Time-dependent SIR model for COVID-19 with undetectable infected persons. Cornell University. doi.org/DOI: https://arxiv.org/abs/2003.00122.

15. Chowell, G., Simonsen, L., Flores, J., Miller, M. A., Viboud, C. (2014). Death patterns during the 1918 influenza pandemic in Chile. Emerging infectious diseases, 20(11), 1803–1811. doi.org/DOI: https://doi.org/10.3201/eid2011.130632

16. Chowell, G., Sattenspiel, L., Bansal, S., Viboud, C. (2016). Mathematical models to characterize early epidemic growth: A review. Physics of life reviews, *18*, 66–97.

17. Elias, C., Nkengasong, J. N., Qadri, F. (2021). Emerging Infectious Diseases - Learning from the Past and Looking to the Future. The New England journal of medicine, 384(13), 1181–1184. doi.org/DOI: https://doi.org/10.1056/NEJMp2034517

18. Diekmann, O., Heesterbeek, J.A.P. (2000). Mathematical epidemiology of infectious diseases: model building, analysis and interpretation. (Wiley series in mathematical and computational biology). John Wiley and Sons.

19. Dietz, K. (1993). The estimation of the basic reproduction number for infectious diseases. Statistical methods in medical research, 2(1), 23–41. doi.ord/DOI: https://doi.org/10.1177/096228029300200103

20. Elias, C., Nkengasong, J. N., Qadri, F. (2021). Emerging Infectious Diseases - Learning from the Past and Looking to the Future. The New England journal of medicine, 384(13), 1181–1184. doi.org/DOI: https://doi.org/10.1056/NEJMp2034517

21. GitHub. (2021). Covid19-timeseries. doi.org/DOI: https://github.com/ulklc/covid19-timeseries

22. Gibbs, Josiah Willard (1902). Elementary Principles in Statistical Mechanics. New York: Charles Scribner's Sons.

23. Global COVID-19 Tracker and Interactive Charts. (2021). doi.org/DOI: https://coronavirus.1point3acres.com/en

24. Grassly, N. C., Fraser, C. (2008). Mathematical models of infectious disease transmission. Nature reviews. Microbiology, 6(6), 477–487. doi.org/DOI: https://doi.org/10.1038/nrmicro1845

25. Hall, I. M., Gani, R., Hughes, H. E., Leach, S. (2007). Real-time epidemic forecasting for pandemic influenza. Epidemiology and infection, 135(3), 372–385. doi.org/DOI: https://doi.org/10.1017/S0950268806007084

26. Heffernan, J. M., Smith, R. J., Wahl, L. M. (2005). Perspectives on the basic reproductive ratio. Journal of the Royal Society, Interface, 2(4), 281–293. https://doi.org/10.1098/rsif.2005.0042

27. Herbert, H.W. (2000). The mathematics of infectious diseases. SIAM Review, 42(4), 599–653. doi.org/DOI: https://epubs.siam.org/doi/ref/10.1137/S0036144500371907

28. Hochberg, M.E. (1991). Non-linear transmission rates and the dynamics of infectious disease. Journal of theoretical biology, 153(3), 301–321. https://doi.org/10.1016/s0022-5193(05)80572-7

29. Hong, H.G., Li, Y. (2020). Estimation of time-varying reproduction numbers underlying epidemiological processes: A new statistical tool for the COVID-19 pandemic. PLoS ONE 15(7): e0236464. doi.org/DOI: https://doi.org/10.1371/journal.pone.0236464

30. Hong, H.G. (2021). Time-Varying SIR-based Poisson Model for COVID19. doi.org/DOI: https://younghhk.shinyapps.io/tvSIRforCOVID19/

31. Horst R.,T. (2018). Mathematics in Population Biology. Princeton University Press. doi.org/DOI: https://doi.org/10.2307/j.ctv301f9v

32. IHME.(2021). doi.org/DOI: http://www.healthdata.org/

33. Jeon, K., Kang, C. I., Yoon, C. H., Lee, D. J., Kim, C. H., Chung, Y. S., Kang, C., Choi, C. M. (2007). High isolation rate of adenovirus serotype 7 from South Korean military recruits with mild acute respiratory disease. European journal of clinical microbiology & infectious diseases: official publication of the European Society of Clinical Microbiology, 26(7), 481–483. doi.org/DOI: https://doi.org/10.1007/s10096-007-0312-6

34. Johns Hopkins Corona virus Resource Center. (2020). https://coronavirus.jhu.edu/

35. Kaul, D.R, Anesi, G.L., et al. (2021). COVID-19–Associated Mortality in Critically Ill Patients Declined During the Pandemic's Early Surge. Journal Watch. doi.org/DOI: https://www.jwatch.org/na53113/2021/02/03/covid-19-associated-mortality-critically-ill-patients

36. Keeling, M., Rohani. P. (2008). Modeling Infectious Diseases: In Humans and Animals. Princeton University Press. doi.org/DOI: https://doi.org/10.2307/j.ctvcm4gk0

37. Kelso, J. K., Milne, G. J., Kelly, H. (2009). Simulation suggests that rapid activation of social distancing can arrest epidemic development due to a novel strain of influenza. BMC public health, 9, 117. doi.org/DOI: https://doi.org/10.1186/1471-2458-9-117

38. Kermack, W.O., McKendrick, A.G. (1927). A contribution to the mathematical theory of epidemics. Proceedings of the Royal Society of London. A 115: 700–721. doi.org/DOI: https://doi.org/10.1098/rspa.1927.0118

39. Matthew Richey (2010). The Evolution of Markov Chain Monte Carlo Methods, The American Mathematical Monthly, 117:5, 383-413, DOI: 10.4169/000298910X485923

40. Mc Kendrick, A.G.(1926). Applications of mathematics to medical problems. Proc. Edinburg Math. Society, 14, 98-130.

41. Kribs-Zaleta, C. M., Velasco-Hernández, J. X. (2000). A simple vaccination model with multiple endemic states. Mathematical biosciences, 164(2), 183–201. doi.org/DOI: https://doi.org/10.1016/s0025-5564(00)00003-1

42. Kroese, D. P., Brereton, T., Taimre, T., Botev, Z. I. (2014). Why the Monte Carlo method is so important today. WIREs Comput Stat. **6** (6): 386–392.

43. Lee, B. Y., Brown, S. T., Cooley, P. C., Zimmerman, R. K., Wheaton, W. D., Zimmer, S. M., Grefenstette, J. J., Assi, T. M., Furphy, T. J., Wagener, D. K., Burke, D. S. (2010). A computer simulation of employee vaccination to mitigate an influenza epidemic. American journal of preventive medicine, 38(3), 247–257. doi.org/DOI: https://doi.org/10.1016/j.amepre.2009.11.009

44. Lessler, J., Reich, N. G., Cummings, D. A., New York City Department of Health and Mental Hygiene Swine Influenza Investigation Team, Nair, H. P., Jordan, H. T., Thompson, N. (2009). Outbreak of 2009 pandemic influenza A (H1N1) at a New York City school. The New England journal of medicine, 361(27), 2628–2636. doi.org/DOI: https://doi.org/10.1056/NEJMoa0906089

45. Miles (Jr.), E.P. (1960). Generalized Fibonacci Numbers and Associated Matrices, The American Mathematical Monthly, 67:8,745-75.

46. Mostafa Adimy, Abdennasser Chekroun, Toshikazu Kuniya. (2022). Traveling waves of a differential-difference diffusive Kermack-McKendrick epidemic model with age-structured protection phase, Journal of Mathematical Analysis and Applications,505(1).

47. Murray, C. (2020). Forecasting COVID-19 impact on hospital bed-days: ICU-days, ventilator days and deaths by US state in the next 4 months. MedRxiv. https://doi.org/10.1101/2020.03.27.20043752

48. Richard T. E. III. (2021). Our Recent Coverage of COVID-19. New England J. Med. doi.org/DOI: https://www.jwatch.org/na53439/2021/03/31/our-recent-coverage-covid-19

49. Ross, R. (1910). The Prevention of Malaria. New York, Dutton.

50. Rubinstein, Reuven Y. (1981). Simulation and the Monte Carlo Method, MAA Review, Wiley Series in Probability and Statistics, Wiley & Sons.

51. Russell, T.W., Hellewell, J., Abbott, S., Jarvis, C.I., Van Zandvoort, K., et al. (2020). Using a delay-adjusted case fatality ratio to estimate under-reporting. Centre for Mathematical Modeling of Infectious Diseases Repository. doi.org/DOI: https://cmmid.github.io/topics/covid19/severity/global_cfr_estimates.html

52. Sanche, S., Lin, Y. T., Xu, C., Romero-Severson, E., Hengartner, N., Ke, R. (2020). High Contagiousness and Rapid Spread of Severe Acute Respiratory Syndrome Coronavirus 2. Emerging infectious diseases, 26(7), 1470–1477. https://doi.org/10.3201/eid2607.200282

53. Strogatz, S. H. (2014). Nonlinear dynamics and chaos: With applications to physics, biology, chemistry, and engineering. 6000 Broken Sound Parkway NW, Suite 300 Boca Raton, FL: Westview press.

54. Tkachenko, A. V., Maslov, S., Elbanna, A., Wong, G. N., Weiner, Z. J., Goldenfeld, N. (2021). Time-dependent heterogeneity leads to transient suppression of the COVID-19 epidemic, not herd immunity. Proceedings of the National Academy of Sciences of the United States of America, 118(17), e2015972118. doi.org/DOI: https://doi.org/10.1073/pnas.2015972118

55. Van den Driessche, P., Watmough, J. (2000). A simple sis epidemic model with a backward bifurcation. Journal of Mathematical Biology, 40(6), 525–540. doi.org/DOI: https://www.meta.org/papers/a-simple-sis-epidemic-model-with-a-backward/10945647

56. Van den Driessche, P., Watmough, J. (2002). Reproduction numbers and sub-threshold endemic equilibria for compartmental models of disease transmission. Mathematical biosciences, 180, 29–48. doi.org/DOI: https://doi.org/10.1016/s0025-5564(02)00108-6

57. Vynnycky, E., White, R. (2010). An introduction to infectious disease modelling. Oxford: Oxford University Press.

58. Wang, C., Liu, L., Hao, X., Guo, H., Wang, Q., Huang, J., et al. (2020). Evolving epidemiology and impact of non-pharmaceutical interventions on the outbreak of coronavirus disease 2019 in Wuhan, China. medRxiv. https://doi.org/10.1101/2020.03.03.20030593

59. Wang, W., Zhao, X-Q. (2008). Threshold dynamics for compartmental epidemic models in periodic environments, J. Dynamics and Differential Equations, 20(3), 699–717.

60. Wesley, C. L., Allen, L. J. S. (2009). The basic reproduction number in epidemic models with periodic demographics. Journal of Biological Dynamics, 3(2–3), 116–129.

61. World Health Organization. (2021). Coronavirus disease 2019 (COVID-19) situation report–31. https://www.who.int/docs/default-source/coronaviruse/situation-reports/20200220-sitrep-31-covid-19.pdf

62. World Health Organization. Evolution of a pandemic (2010). A (H1N1) 2009, April 2009–March 2010. Retrieved 9 July 2019 from https://www.who.int/csr/disease/swineflu/laboratory1_04_2010/en/

63. Yang, X., Wang Q., Liang, B., Wu, F., Li, H., Liu, H., et al. (2015). An outbreak of acute respiratory disease caused by a virus associated RNA II gene mutation strain of human adenovirus 7 in China, PLOS One.
 https://journals.plos.org/plosone/article?id=10.1371/journal.pone.0172519

64. Yang, Z., Zhou, T. (2012). Epidemic spreading in weighted networks: An edge-based mean-field solution, Phys. Rev. E 85, 056106.

65. Yu P., Ma C., Nawaz M., Han L., Zhang J., Du Q., Zhang L., Feng Q., Wang J., Xu J. (2013). Outbreak of acute respiratory disease caused by human adenovirus type 7 in a military training camp in Shaanxi, China. Microbiology and immunology, 57(8): 553–560.

66. Yusof, M., Rashid, T., Thayan, R., Othman, K., Abu, Hasan, N., Adnan, N., Saat, Z. (2012). Human Adenovirus Type 7 Outbreak in Police Training Center, Malaysia, 2011. Emerging Infectious Diseases, 18(5): 852-854.

67. Zhao, S., Wan, Cheng-Song, et al. (2014). Re-emergent human adenovirus genome type 7d caused an acute respiratory disease outbreak in Southern China after a twenty-one-year absence. Scientific reports. 4: 7365.